\documentclass[aps,prl,reprint,groupedaddress]{revtex4-1}
\usepackage{amssymb}
\usepackage{amsmath}
\usepackage{bbold}
\usepackage{slashed}
\usepackage{IEEEtrantools}
\usepackage{physics}
\usepackage{bm}
\usepackage{longtable}

\newcommand{\ud}{\mathrm{d}}
\begin{document}

% Use the \preprint command to place your local institutional report
% number in the upper righthand corner of the title page in preprint mode.
% Multiple \preprint commands are allowed.
% Use the 'preprintnumbers' class option to override journal defaults
% to display numbers if necessary
%\preprint{}

%Title of paper
\title{Mass measurable phase differences in neutrino oscillations}

% repeat the \author .. \affiliation  etc. as needed
% \email, \thanks, \homepage, \altaffiliation all apply to the current
% author. Explanatory text should go in the []'s, actual e-mail
% address or url should go in the {}'s for \email and \homepage.
% Please use the appropriate macro foreach each type of information

% \affiliation command applies to all authors since the last
% \affiliation command. The \affiliation command should follow the
% other information
% \affiliation can be followed by \email, \homepage, \thanks as well.
%\author{Xiang Zhou}
\author{Xiang Zhou} %(周详)
\email[]{xiangzhou@whu.edu.cn}
%\homepage[]{Your web page}
%\thanks{}
%\altaffiliation{}
\affiliation{Hubei Nuclear Solid Physics Key Laboratory, School of Physics and Technology, Wuhan University, Wuhan 430072, China}

%Collaboration name if desired (requires use of superscriptaddress
%option in \documentclass). \noaffiliation is required (may also be
%used with the \author command).
%\collaboration can be followed by \email, \homepage, \thanks as well.
%\collaboration{}
%\noaffiliation

\date{\today}

\begin{abstract}
% insert abstract here
	We propose an ansatz, named the ``particle ansatz'', that in the rest frame of a flavor neutrino all its mass eigenstate momentums should simultaneously be zero. Under the particle ansatz, the mass measurable phase differences (mmPhDs) can be derived as $\Delta\phi_{ji}\simeq [2m_{\nu_\alpha}(m_j-m_i)/4 E_{\nu_\alpha}]L$. We show that using mmPhDs all three neutrino masses can be determined. In the foreseeable future all neutrino masses are expected to be determined by the spectral analysis of neutrino experiments.  
\end{abstract}

% insert suggested PACS numbers in braces on next line
% insert suggested keywords - APS authors don't need to do this
%\keywords{}

%\maketitle must follow title, authors, abstract, \pacs, and \keywords
\maketitle

% body of paper here - Use proper section commands
% References should be done using the \cite, \ref, and \label commands
%\section{}
% Put \label in argument of \section for cross-referencing
%\section{\label{}}
%\subsection{}
%\subsubsection{}
It is well known that neutrinos have nonzero masses because of neutrino oscillations. However, it is believed that neutrino oscillation experiments can not determine the absolute neutrino masses but the mass square difference $\Delta m_{ji}^2=m_j^2-m_i^2$. The origin of this faith is based on the phase difference (PhD) in the traditional phenomenological theory of neutrino oscillations which is 
\begin{equation}
	\Delta\phi_{ji}\simeq\frac{m_j^2-m_i^2}{4E_\nu}L=\frac{\Delta m_{ji}^2}{4E_\nu}L,
	\label{oldPhD}
\end{equation}
where $m_i$ and $m_j$ are the mass eigenvalues, $E_\nu$ is the neutrino energy and L is the neutrino flight distance. PhD in Eq.~(\ref{oldPhD}) can be derived under the so called the ``same energy'' or ``same momentum'' ansatz\cite{Thomson}. However, either the ``same energy'' or ``same momentum'' ansatz has been shown to contradict Lorentz invariance and lead to a number of paradoxes\cite{Akhmedov09,Akhmedov11}. In this paper, an ansatz is proposed from which a new form of phase difference $\Delta\phi_{ji}\simeq [2m_{\nu_\alpha}(m_j-m_i)/4 E_{\nu_\alpha}]L$ can be derived. We show that the neutrino masses can in principle be determined.  

The Dirac equation for a spin $1/2$ particle with mass $m_i$ is $(i\gamma^\mu\partial_\mu-m_i)\psi_i=0$. A set of orthogonal mass eigenstates can be defined as $\ket{\nu_i}=(0,\ldots,\psi_i,\ldots,0)^T$ with $\bra{\nu_j}\ket{\nu_i}=\delta_{ij}$ and $i,j=1,2,\ldots,n$. There can be at most $n$ linear independent combinations which are written as
\begin{equation}
	\ket{\nu_\alpha}=\sum_{i} U^*_{\alpha i}\ket{\nu_i},
\end{equation}
where $\alpha=\alpha_1,\alpha_2,\ldots,\alpha_n$. If $\bra{\nu_\beta}\ket{\nu_\alpha}=\delta_{\beta\alpha}$, the matrix $U^*=[U^*_{\alpha i}]$ is a unitary one. Neutrino oscillation shows that a flavor neutrino $\nu_\alpha$ has nonzero mass and its flavor eigenstates $\ket{\nu_\alpha}$ is different to the mass eigenstates $\ket{\nu_i}$. Thus, $U^*$ has nonzero off-diagonal elements. 

Since a flavor neutrino experimentally behaves as a particle, there should be a rest frame for it where its momentum is zero. We propose an ansatz called the ``particle ansatz'' that in the rest frame of a flavor neutrino $\nu_\alpha$ all momentums of its mass eigenstates $\ket{\nu_i}$ are simultaneously zero so that  
\begin{equation}
	\ket{\nu_\alpha(\tau)}=\sum_{i} U^*_{\alpha i}\ket{\nu_i(\tau)}=\sum_{i} U^*_{\alpha i}\ket{\nu_i(\tau_0)}e^{-im_i\tau},
\end{equation}
where $\tau$ is the proper time, $\tau_0=0$ is the initial proper time and $\ket{\nu_i(\tau_0)}$ is the zero-momentum wavefunction in the rest frame. In the lab frame $\tau$ is boosted to $t$ and $t_0=0$ so that the wavefunction of $\nu_\alpha$ should be 
\begin{eqnarray}
	\ket{\nu_\alpha(t)} & = & \sum_{i} U^*_{\alpha i}\ket{\nu_i(t)} \nonumber\\
	                    & = & \sum_{i} U^*_{\alpha i}\ket{\nu_i(t_0)}e^{-i(E_it-\mathbf{p}_i\cdot\mathbf{x})},
\end{eqnarray}
where $E_i$ and $\mathbf{p}_i$ are the energy and momentum of the mass eigenstate $\ket{\nu_i(t)}$ with $E_i=\gamma m_i$ and $\gamma=\mathrm{d}\tau/\mathrm{d}t$. Thus the the particle ansatz can induce the ``same velocity'' condition\cite{Thomson,Suekane} but velocity is not a well-defined quantum mechanical concept because of Heisenberg's uncertainty principle.

The mass and mass square of $\nu_\alpha$ can be obtained by
\begin{eqnarray}
	m_{\nu_\alpha}   & = & \bra{\nu_\alpha(\tau)}i\frac{\ud}{\ud \tau}\ket{\nu_\alpha(\tau)} \nonumber\\ 
	                 & = & \sum_{j}\sum_{i}\bra{\nu_j(\tau)}U_{\beta j}m_iU^*_{\alpha i}\ket{\nu_i(\tau)} \nonumber\\
			 & = & \sum_{i} m_i|U_{\alpha i}|^2, 
\end{eqnarray}
\begin{eqnarray}
	m_{\nu_\alpha}^2 & = & \bra{\nu_\alpha(\tau)}-\frac{\ud^2}{\ud \tau^2}\ket{\nu_\alpha(\tau)} \nonumber\\ 
	                 & = & \sum_{j}\sum_{i}\bra{\nu_j(\tau)}U_{\beta j}m_i^2U^*_{\alpha i}\ket{\nu_i(\tau)} \nonumber\\
			 & = & \sum_{i} m_i^2|U_{\alpha i}|^2
\end{eqnarray}
The energy and energy square of $\nu_\alpha$ in the lab frame can be obtained by
\begin{eqnarray}
	E_{\nu_\alpha}   & = & \bra{\nu_\alpha(t)}i\frac{\ud}{\ud t}\ket{\nu_\alpha(t)} \nonumber\\
	                 & = & \sum_{j}\sum_{i}\bra{\nu_j(t)}U_{\beta j}E_iU^*_{\alpha i}\ket{\nu_i(t)} \nonumber\\
			 & = & \sum_{i} E_i|U_{\alpha i}|^2, 
\end{eqnarray}
\begin{eqnarray}
	E_{\nu_\alpha}^2 & = & \bra{\nu_\alpha(t)}-\frac{\ud^2}{\ud t^2}\ket{\nu_\alpha(t)} \nonumber\\
	                 & = & \sum_{j}\sum_{i}\bra{\nu_j(t)}U_{\beta j}E_i^2U^*_{\alpha i}\ket{\nu_i(t)} \nonumber\\
			 & = & \sum_{i} E_i^2|U_{\alpha i}|^2.
\end{eqnarray}
The momentum and momentum square of $\nu_\alpha$ in the lab frame can be obtained by
\begin{eqnarray}
	\mathbf{p}_{\nu_\alpha}   & = & \bra{\nu_\alpha(t)}-i\frac{\ud}{\ud \mathbf{x}}\ket{\nu_\alpha(t)} \nonumber\\
	                          & = & \sum_{j}\sum_{i}\bra{\nu_j(t)}U_{\beta j}\mathbf{p}_iU^*_{\alpha i}\ket{\nu_i(t)} \nonumber\\
				  & = & \sum_{i}\mathbf{p}_i|U_{\alpha i}|^2, 
\end{eqnarray}
\begin{eqnarray}
	\mathbf{p}_{\nu_\alpha}^2 & = & \bra{\nu_\alpha(t)}-\frac{\ud^2}{\ud \mathbf{x}^2}\ket{\nu_\alpha(t)} \nonumber\\
	                          & = & \sum_{j}\sum_{i}\bra{\nu_j(t)}U_{\beta j}\mathbf{p}_i^2U^*_{\alpha i}\ket{\nu_i(t)} \nonumber\\
				  & = & \sum_{i} \mathbf{p}_i^2|U_{\alpha i}|^2, 
\end{eqnarray}
Since $E_i^2-\mathbf{p}_i^2=m_i^2$ and $E_i=\gamma m_i$, $E_{\nu_\alpha}^2-\mathbf{p}_{\nu_\alpha}^2=m_{\nu_\alpha}^2$ and $E_{\nu_\alpha}=\gamma m_{\nu_\alpha}$. Therefore, a flavor neutrino $\nu_\alpha$ described by $\ket{\nu_\alpha(t)}$ does behave like a relativistic particle.

The flavor wavefunctions $\ket{\nu_\alpha(\tau_0)}$ or $\ket{\nu_\alpha(t_0)}$ with $\alpha=\alpha_1,\ldots,\alpha_n$ can also be chosen as a set of bases for $\nu_\alpha(\tau)$ or $\ket{\nu_\alpha(t)}$, respectively so that in the rest frame\begin{eqnarray}
	\ket{\nu_\alpha(\tau)} & = & \sum_{\beta}\sum_i|U_{\beta i}|^2e^{-im_i\tau}\ket{\nu_\beta(\tau_0)} \nonumber \\
	                       & = & \sum_{\beta}C_\beta(\tau)\ket{\nu_\beta(\tau_0)},
\end{eqnarray}
and in the lab frame
\begin{eqnarray}
	\ket{\nu_\alpha(t)} & = & \sum_{\beta}\sum_i|U_{\beta i}|^2e^{-i(E_it-\mathbf{p}_i\cdot\mathbf{x})}\ket{\nu_\beta(t_0)} \nonumber \\
                            & = & \sum_{\alpha}C_\beta(t)\ket{\nu_\beta(t_0)}. 
			    \end{eqnarray}
Since the phase of $\ket{\nu_i}$ satisfies $\phi_i=m_i\tau=E_it-\mathbf{p}\cdot\mathbf{x}$, the coefficients $C_\beta(\tau)$ and $C_\beta(t)$ are Lorentz invariant and satisfy 
\begin{equation}
	C_\beta=C_\beta(\tau)=C_\beta(t)=\sum_i|U_{\beta i}|^2e^{-i\phi_i}.
\end{equation}
Thus the probability of observing $\ket{\nu_\beta}$ at $\Delta\tau=\tau-\tau_0$ in the rest frame or $\Delta t=\Delta\tau/\gamma=t-t_0$ in the lab frame is also Lorentz invariant which is
\begin{eqnarray}
	P(\nu_\alpha\to\nu_\beta) & = & |\bra{\nu_\beta(\tau_0)}\ket{\nu_\alpha(\tau)}|^2=|\bra{\nu_\beta(t_0)}\ket{\nu_\alpha(t)}|^2 \nonumber\\
	                          & = & |C_\beta|^2=|\sum_i U_{\beta i}U^*_{\alpha i}e^{-i\phi_i}|^2 \nonumber\\
				  & = & \sum_j\sum_iU_{\alpha j}U^*_{\beta j}U^*_{\alpha i}U_{\beta i}e^{i2\Delta\phi_{ji}} \nonumber\\ 
				  & = & \delta_{\alpha\beta}-4\sum_{j>i}\Re{U_{\alpha j}U^*_{\beta j}U^*_{\alpha i}U_{\beta i}}\sin^2\Delta\phi_{ji} \nonumber\\ 
				  &   & -2\sum_{j>i}\Im{U_{\alpha j}U^*_{\beta j}U^*_{\alpha i}U_{\beta i}}\sin 2\Delta\phi_{ji},
\end{eqnarray}
where the PhD $\Delta\phi_{ji}$ is defined as
\begin{equation}
	\Delta\phi_{ji}=\frac{\phi_j-\phi_i}{2}=\frac{m_j-m_i}{2}\Delta\tau=\frac{m_j-m_i}{2}\frac{\Delta t}{\gamma}.
	\label{particlePhD}
\end{equation}

The arithmetic mean of $E_j$ and $E_i$ is defined as $\overline{E}_{ji}=(E_j+E_i)/2$ and the arithmetic mean of $m_j$ and $m_i$ is defined as $\overline{m}_{ji}=(m_j+m_i)/2$\cite{Suekane}. Because 
\begin{equation}
	\gamma=\frac{\ud\tau}{\ud t}=\frac{E_i}{m_i}=\frac{\overline{E}_{ji}}{\overline{m}_{ji}},
\end{equation}
the PhD in Eq.~(\ref{oldPhD}) derived from the ``same energy'' or ``same momentum'' ansatz can be obtained by
\begin{eqnarray}
	\Delta\phi_{ji} & = & \frac{\overline{m}_{ji}(m_j-m_i)}{2 \overline{E}_{ji}}\Delta t \nonumber\\
	            & = & \frac{m_j^2-m_i^2}{4\overline{E}_{ji}}\Delta t\simeq\frac{m_j^2-m_i^2}{4\overline{E}_{ji}}L,
\end{eqnarray}
where $L$ is the distance and  $\Delta t\simeq L$ is used since $\gamma\gg 1$. 

However $\overline{E}_{ji}$ is not any more a measurable quantity in the particle ansatz. In the particle ansatz the measurable energy of $\nu_\alpha$ is $E_{\nu_\alpha}$. Because $\gamma=E_{\nu_\alpha}/m_{\nu_\alpha}$, PhD in Eq.~(\ref{particlePhD}) can be rewritten as 
\begin{eqnarray}
	\Delta\phi_{ji} & =      & \frac{m_{\nu_\alpha}(m_j-m_i)}{2 E_{\nu_\alpha}}\Delta t \nonumber\\
		    & \simeq & \frac{2m_{\nu_\alpha}(m_j-m_i)}{4 E_{\nu_\alpha}}L=\frac{\Delta m^2_{\alpha ji}}{4 E_{\nu_\alpha}}L,
		    \label{mmPhD}
\end{eqnarray}
where $\Delta m^2_{\alpha ji}=2m_{\nu_\alpha}(m_j-m_i)$. The mass square term $\Delta m^2_{\alpha ji}$ in the above PhD is the product of the double flavor neutrino mass $2m_{\nu_\alpha}$ and the difference of two mass eigenvalues $m_j$ and $m_i$. Because three flavor neutrino masses are generally independent to each other, the neutrino mass can be determined by neutrino oscillation experiments. Therefore the Ph.D in Eq.~(\ref{mmPhD}) is a mass measurable phase difference (mmPhD). 

For the three neutrino oscillations, $m_{\nu_\alpha}$ and $m_i$ have the relationship as
\begin{equation}
	m_{\nu_\alpha}=\sum_i|U_{\alpha i}|^2m_i\quad\mathrm{and}\quad m_i=\sum_\alpha\overline{U^2}_{\alpha i}m_{\nu_\alpha},
\end{equation}
where $i=1,2,3$ and $\alpha=e,\mu,\tau$ and
\begin{equation}
	\sum_{\alpha}|U_{\alpha j}|^2\overline{U^2}_{\alpha i}=\delta_{ji}.
\end{equation}
$U_{\alpha i}$ is the element of the PMNS matrix $U=[U_{\alpha i}]$ parameterized by three mixing angles $\theta_{12}$, $\theta_{23}$ and $\theta_{13}$ and Dirac phase $\delta$ which satisfies
\begin{widetext}
\begin{equation}
	U_\mathrm{PMNS}=
	\begin{pmatrix}
		U_{e 1}    & U_{e 2}    & U_{e 3}    \\
		U_{\mu 1}  & U_{\mu 2}  & U_{\mu 3}  \\
		U_{\tau 1} & U_{\tau 2} & U_{\tau 3} \\
	\end{pmatrix}=\
	\begin{pmatrix}
		1 & 0       & 0      \\
		0 &  c_{23} & s_{23} \\
		0 & -s_{23} & c_{23} \\
	\end{pmatrix}
	\begin{pmatrix}
		c_{13}            & 0 & s_{13}e^{-i\delta} \\
		0                 & 1 & 0                  \\
	       -s_{13}e^{i\delta} & 0 & c_{13}             \\
	\end{pmatrix}
	\begin{pmatrix}
		c_{12} & s_{12} & 0 \\
	       -s_{12} & c_{12} & 0 \\
		0      & 0      & 1 \\
	\end{pmatrix},
\end{equation}
\end{widetext}
where $c_{ij}=\cos\theta_{ij}$ and $s_{ij}=\sin\theta_{ij}$. $\overline{U^2}_{\alpha i}$ is the element of the inverse matrix of $[|U_{\alpha i}|^2]$. If $\delta=\pm 90^\circ$ and $\theta_{23}=45^\circ$ where the determinant of the matrix $[|U_{\alpha i}|^2]$ is zero and then the matrix $[\overline{U^2}_{\alpha i}]$ can not obtained. It can be easily shown that
\begin{equation}
	M=m_{\nu_e}+m_{\nu_\mu}+m_{\nu_\tau}=m_1+m_2+m_3,
\end{equation}
where $M$ is the total mass of neutrinos. If the coefficients $k=m_{\nu_e}/M$ and $\epsilon=m_{\nu_\mu}/m_{\nu_e}$ are introduced, the masses of flavor neutrinos can be written as $m_{\nu_e}=kM$, $m_{\nu_\mu}=\epsilon kM$ and $m_{\nu_\tau}=(1-k-\epsilon k)M$ with $0<k<1$, $0<\epsilon k<1$ and $0<1-k-\epsilon k<1$.

For neutrino experiments the traditional mass square terms in the PhDs under the ``same energy'' or ``same momentum'' can be transformed to the new mass square terms in their corresponding mmPhDs which are listed as
\begin{equation}
\begin{array}{lcllll}
	P(\nu_e\to\nu_e)                 &  \Delta m_{21}^2  & \to & \Delta m^2_{e21}    & = & 2m_{\nu_e}(m_2-m_1)          \nonumber\\
	P(\bar{\nu}_e\to\bar{\nu}_e)	 &                   &     &                     & = & 2m_{\bar{\nu}_e}(m_2-m_1),   \nonumber\\
	    			          		                                                                    \nonumber\\
	P(\nu_\mu\to\nu_\mu)             & |\Delta m_{32}^2| & \to & \Delta m^2_{\mu 32} & = & 2m_{\nu_\mu}|m_3-m_2|        \nonumber\\
	P(\bar{\nu}_\mu\to\bar{\nu}_\mu) &                   &     &                     & = & 2m_{\bar{\nu}_\mu}|m_3-m_2|, \nonumber\\
	    			          		                                                                    \nonumber\\
	P(\nu_e\to\nu_e)                 & |\Delta m_{32}^2| & \to & \Delta m^2_{e32}    & = & 2m_{\nu_e}|m_3-m_2|          \nonumber\\
	P(\bar{\nu}_e\to\bar{\nu}_e)     &                   &     &                     & = & 2m_{\bar{\nu}_e}|m_3-m_2|,   \nonumber\\
\end{array}
\end{equation}
where $\epsilon=m_{\nu_\mu}/m_{\nu_e}=\Delta m^2_{\mu 32}/\Delta m^2_{e32}$ can be obtained in principle. Thus, $\Delta m^2_{e21}$ and $\Delta m^2_{\mu 32}$ can be rewritten as
\begin{eqnarray}
	\Delta m^2_{e21} & = & 2m_{\nu_e}[(\overline{U^2}_{e2}-\overline{U^2}_{e1})m_{\nu_e}               \nonumber\\
			 &   & +(\overline{U^2}_{\mu 2}-\overline{U^2}_{\mu 1})m_{\nu_\mu}                 \nonumber\\
			 &   & +(\overline{U^2}_{\tau 2}|^2-\overline{U^2}_{\tau 1})m_{\nu_\tau}]          \nonumber\\
			 & = & 2[(\overline{U^2}_{e2}-\overline{U^2}_{e1})k^2                              \nonumber\\
			 &   & +(\overline{U^2}_{\mu 2}-\overline{U^2}_{\mu 1})\epsilon k^2                         \\
			 &   & +(\overline{U^2}_{\tau 2}-\overline{U^2}_{\tau 1})(k-k^2-\epsilon k^2)]M^2, \nonumber                                     
		   \label{me21}
\end{eqnarray}
\begin{eqnarray}
	\Delta m^2_{\mu 32} & = & 2m_{\nu_\mu}|(\overline{U^2}_{e3}-\overline{U^2}_{e2})m_{\nu_e}                                 \nonumber\\
			    &   & +(\overline{U^2}_{\mu 3}-\overline{U^2}_{\mu 2})m_{\nu_\mu}                                     \nonumber\\
			    &   & +(\overline{U^2}_{\tau 3}-\overline{U^2}_{\tau 2})m_{\nu_\tau}|                                 \nonumber\\
			    & = & 2|(\overline{U^2}_{e3}-\overline{U^2}_{e2})\epsilon k^2                                         \nonumber\\
			    &   & +(\overline{U^2}_{\mu 3}-\overline{U^2}_{\mu 2})\epsilon^2k^2                                            \\
			    &   & +(\overline{U^2}_{\tau 3}-\overline{U^2}_{\tau 2})(\epsilon k-\epsilon k^2-\epsilon^2 k^2)|M^2. \nonumber
		   \label{mmu32}
\end{eqnarray}

If the parameters of PMNS matrix are known and $\epsilon=m_{\nu_\mu}/m_{\nu_e}$ can be obtained by measuring the ratio of $\Delta m_{\mu 32}$ and $\Delta m_{e 32}$ from neutrino experiments, the two free parameters, the total mass of neutrinos $M$ and the mass ratio $k=m_{\nu_e}/M$, can be calculated from Eq.~(\ref{me21}) and Eq.~(\ref{mmu32}) and then all masses of flavor and mass eigenstate neutrinos can be obtained. $\epsilon$ is around 1 because there is not much difference between the values of $|\Delta m_{32}^2|=|m_3^2-m_2^2|$ measured by electron and muon neutrino oscillation experiments\cite{KamLAND2005,MINOS2014,DayaBay2017,T2K2017,NOVA2017,PDG2016}. Table~\ref{mass} gives the calculated results where $\epsilon$ runs from 0.90 to 1.10, and the input neutrino oscillation parameters are
\begin{equation}
	\begin{array}{c}
		\theta_{23}\sim 40^\circ~\mathrm{or}~50^\circ \\
	\theta_{12}\sim 35^\circ, \quad \theta_{13}\sim 9^\circ, \quad \delta\sim \pm 90^\circ \\
		\Delta m^2_{e21}\sim 8 \times 10^{-5} \mathrm{eV}^2, \quad \Delta m^2_{\mu 32}\sim 2.5 \times 10^{-3} \mathrm{eV}^2. 
	\end{array}
	\label{inputs}
\end{equation}
\begin{table}[b]
	\caption{\label{mass}The calculated total mass of neutrinos $M$ and neutrino mass hierarchy (MH) with the input neutrino oscillation parameters in Eq.~(\ref{inputs}). NH is the normal hierarchy and IH is inverted hierarchy.}
\begin{ruledtabular}
\begin{tabular}{cccccc}
$\theta_{12}$ & $\theta_{23}$ & $\theta_{13}$ & $\delta$       & $\Delta m^2_{e21}$ (eV$^2$) & $\Delta m^2_{\mu 32}$ (eV$^2$) \\
$35^\circ$    & $40^\circ$    & $9^\circ$     & $\pm 90^\circ$ & $8 \times 10^{-5}$          & $2.5 \times 10^{-3}$           \\
\colrule
	      & $\epsilon$    & $k$           & MH             & $M$ (eV)         \\                    
              & 1.10          & 0.308         & NH             & 0.217            \\ 
              & 1.08          & 0.313         & NH             & 0.241            \\ 
              & 1.06          & 0.318         & NH             & 0.277            \\ 
              & 1.04          & 0.323         & NH             & 0.337            \\ 
              & 1.02          & 0.328         & NH             & 0.473            \\ 
              & 1             & 1/3           & N/A            & $\infty$         \\ 
	      & 0.98          & 0.339         & IH             & 0.450            \\
	      & 0.96          & 0.345         & IH             & 0.316            \\
	      & 0.94          & 0.351         & IH             & 0.257            \\
	      & 0.92          & 0.357         & IH             & 0.221            \\
	      & 0.90          & 0.363         & IH             & 0.196            \\
\colrule
$\theta_{12}$ & $\theta_{23}$ & $\theta_{13}$ & $\delta$       & $\Delta m^2_{e21}$ (eV$^2$) & $\Delta m^2_{\mu 32}$ (eV$^2$) \\
$35^\circ$    & $50^\circ$    & $9^\circ$     & $\pm 90^\circ$ & $8 \times 10^{-5}$          & $2.5 \times 10^{-3}$           \\
\colrule
	      & $\epsilon$    & $k$           & MH             & $M$ (eV)    \\                    
	      & 1.10          & 0.315         & NH             & 0.254       \\ 
	      & 1.08          & 0.319         & NH             & 0.284       \\ 
	      & 1.06          & 0.322         & NH             & 0.327       \\ 
	      & 1.04          & 0.326         & NH             & 0.400       \\ 
	      & 1.02          & 0.330         & NH             & 0.564       \\ 
	      & 1             & 1/3           & N/A            & $\infty$    \\ 
              & 0.98          & 0.337         & IH             & 0.546       \\   
              & 0.96          & 0.341         & IH             & 0.386       \\   
              & 0.94          & 0.345         & IH             & 0.315       \\   
              & 0.92          & 0.349         & IH             & 0.273       \\   
              & 0.90          & 0.353         & IH             & 0.244       \\   
\end{tabular}
\end{ruledtabular}
\end{table}

Table~\ref{mass} shows that there are a mass hierarchy (MH) transition around $\epsilon = 1$. Whatever $\theta_{23}\sim 40^\circ$ or $50^\circ$, MH is normal when $\epsilon>1$ but inverted when $\epsilon<1$. The total masses of neutrinos $M$ depends on the octant of $\theta_{23}$. Except for the vicinity of MH transition point, $M\sim O(10^{-1})$ eV which coincides with the constraints from cosmological and astrophysical models\cite{PDG2016}. Therefore the cosmological and astrophysical models can constrain in return the parameters of neutrino oscillations, such as, Dirac phase $\delta$ and octant of $\theta_{23}$. The general solutions of Eq.~(\ref{me21}) and Eq.~(\ref{mmu32}) will be discussed in Ref.~\cite{Wu}.

We have shown that the particle ansatz induces the mmPhDs with which the neutrino masses can be determined in principle. The masses of all three flavor neutrinos can not be the same. Because three mmPhDs are mixed in neutrino oscillations, the precise value of $\epsilon=m_{\nu_\mu}/m_{\nu_e}=\Delta m^2_{\mu 32}/\Delta m^2_{e32}$ has to be determined by the spectral analysis of neutrino experiments under the particle ansatz which is beyond the scope of this paper. It is expected that with the mixing angles, Dirac phase, and mass hierarchy all neutrino masses can be also experimentally determined in the foreseeable future\cite{JUNO2016,PDG2016}.

% If you have acknowledgments, this puts in the proper section head.
\begin{acknowledgments}
	I sincerely thank Dr. Zhenyu Zhang's great patience and constructive discussions. Happy birthday to you, my dear mm Ph.D! This work is supported by the Major Program of the National Natural Science Foundation of China (Grant No. 11390381).
\end{acknowledgments}

% Create the reference section using BibTeX:
%\bibliography{basename of .bib file}

\end{document}